\documentstyle[aps,preprint,epsf]{revtex}
\tightenlines
\newcommand{\be}{\begin{eqnarray}}
\newcommand{\ee}{\end{eqnarray}}
\newcommand\del{\partial}

\newcommand\Dels{D \hspace{-0.26cm} /}

\begin{document}

\title{Chiral Symmetry in Two--Color QCD at Finite Temperature} 
\author{J. Wirstam}

\address{
Institute of Theoretical Physics \\  Stockholm University, Box 6730, S-113 85 Stockholm, Sweden \\ 
email: wirstam@physto.se \\ }

\maketitle

\begin{abstract}
We study the chiral symmetry in two--color QCD with $N_f$ massless flavors at finite temperature,
using an effective theory.
For the gauge group SU(2), the chiral symmetry group is enlarged to
SU(2$N_f$), which is then spontaneously
broken to Sp(2$N_f$) at zero temperature. At finite temperature, and when the axial anomaly can be 
neglected, we find a first order phase transition occurring for two or more flavors.
In the presence of instantons, the symmetry restoration unambiguously remains first order for 
three or more massless flavors. These results could be relevant for lattice studies of chiral symmetry
at finite temperature and density.
\end{abstract}

\section{Introduction}

An important issue in QCD is the behavior of the theory
under extreme conditions, such as a high temperature or large density. The high temperature
region is relevant for an understanding of the early stages after the big bang, whereas high densities
occur in e.g. neutron stars. In particular, with the upcoming events at RHIC  
such conditions may be obtained experimentally, thus providing a direct 
test of the theoretical predictions and increasing our knowledge about QCD at high temperatures and densities. 

Of special interest is the phase diagram of QCD, i.e. whether there are any phase transitions or 
cross--over regions when the temperature and chemical potential is increased from zero.
For instance, QCD
is well known to have a phase transition separating a zero temperature phase, where the chiral symmetry is broken, from
a high temperature chirally symmetric phase.
At zero chemical potential, the order of the finite temperature phase transition 
was studied in \cite{Robwilczek}, using the linear $\sigma$--model as an effective theory.
Numerical studies on the lattice agree with the analytical results and indicate a second order
phase transition for N$_f=2$ massless flavors, which becomes first order for N$_f=3$ massless or
light flavors \cite{iwasaki}. The best quantitative estimate for the critical temperature $T_c$ comes from the lattice 
simulations, giving a numerical value around $T_c \simeq 150$ MeV.

Despite the time consuming algorithms for simulating full QCD on the lattice, the finite temperature
observables are at least in principle calculable. However, when it comes to 
the introduction of a chemical potential, $\mu$,
the numerical simulations encounter severe problems. The principal difficulty is due to the functional measure
(or more precisely, the determinant remaining after the fermions have been integrated out), which
becomes complex--valued for finite $\mu$ and  
thus makes the usual importance sampling impossible. Although several different proposals 
have been made in order to overcome this problem, such as the use 
Langevin algorithms \cite{langevin} and the so called Glasgow method \cite{glasgow}, 
the successes have so far been rather limited. A quantitative understanding
of the chiral symmetry restoration in  QCD at finite chemical potential is still lacking.

Considering the above mentioned difficulties, it may very well prove worthwhile to study the finite
chemical potential problem in a somewhat simpler context. In this respect, QCD with SU(2) 
replacing SU(3) as the gauge group is of considerable interest as a toy model. 
Specifically, since the four--dimensional, non--abelian structure is still maintained, some of the general
properties in the simplified theory should be rather similar to the SU(3) theory. Thus,  
with such a minor modification 
one could hopefully obtain some insights of a qualitative nature into the structure of full QCD.

However, although two--color QCD apparently resembles real QCD in many respects, there are nevertheless some striking
differences. Obviously, hadron spectroscopy is going to be completely different. Another example is 
the consequence of a formation of diquark condensates, which has a different interpretation in two--color QCD 
and full QCD. 
In real QCD, a nonvanishing diquark condensate breaks the local gauge symmetry, 
and becomes an order parameter for color superconductivity \cite{bailinlove}, a new state in QCD
which has received much attention lately (see e.g. \cite{wilczek} and references therein).
In two-color QCD, the situation is slightly different since the diquarks 
may form a gauge singlet condensate. The condensate
would still break baryon number conservation though, and hence become an order parameter for superfluidity instead.

The main motivation for replacing SU(3) by SU(2) is the possibility to study the theory numerically. Since the  
fermions are in a pseudo--real representation for SU(2), the functional measure is real and positive,
even with the introduction of a chemical potential. Therefore, two--color QCD at finite density can be 
studied on the lattice using standard methods.
Preliminary lattice studies at finite chemical potential, 
both in the quenched approximation \cite{lattice1}
and with unquenched fermions \cite{lattice2}, indicate a nonvanishing diquark condensate
at any chemical potential $\mu > 0$ in the chiral limit. 
Due to the pseudoreal representation, when $\mu =0$
the $\langle qq \rangle$ and the usual $\langle \bar{q} q \rangle$ condensates can in fact be rotated into each other
without any cost in energy. 
Moreover, analytical results using QCD inequalities \cite{correlators} and models based on instanton
induced interactions \cite{instantons} agree with the overall picture emerging from the numerical results. 

There is thus an increasing amount of evidence in support of the above mentioned behavior of SU(2) QCD at finite $\mu$.
However, very little is still known about the thermodynamics of two--color QCD, 
although some results were obtained rather recently \cite{lombardo}. 
This paper is devoted to such a study of the chiral condensate at finite temperature, 
using an effective theory for the condensate, and the fluctuations around it, to describe the critical behavior.
Our motivation is twofold; first of all, it is theoretically interesting
to investigate the whole phase diagram of two--color QCD, for instance when it comes to the question of
possible phase transitions and the connection to a superfluid phase. Secondly,
it is of course also valuable to have a comparison between 
future numerical efforts in two--color QCD
at finite temperature (and density), 
and other approaches based on different methods. Hopefully the combined methods will give a better understanding
of two--color QCD, as well as a glimpse of the general features in full QCD.

The paper is organized as follows. In the next section we recall the chiral symmetry and 
the symmetry breaking pattern for two--color QCD, and write down an effective theory.
Then in section III we use the effective theory to analyze the chiral symmetry breaking
at finite temperature. We end in section IV by giving our conclusions and an outlook.
   
\section{The Chiral Symmetry in Two--Color QCD}
\label{II}

From now on, we assume that the gauge group is  SU(2) and furthermore that there are $N_f$ flavors of massless quarks.
The part of the QCD action containing the fermions is given by,
\be
S= -i \sum_{f=1}^{N_f} \int d^4 \! x \, \bar{\psi}_f^i \left ( \del_{\mu} \delta^{j}_i -ig A_{\mu}^a (T^a)^{j}_i \right ) 
\gamma^{\mu} \psi_{jf} = -i \sum_{f=1}^{N_f} \int d^4 \! x \, \bar{\psi}_f \Dels \psi_f \ , \label{startaction}
\ee
where the metric is $g_{\mu \nu} = {\rm diag} (+---)$ and the
color matrices $T^a = \sigma^a /2$, with $\sigma^a$ the usual Pauli matrices.
Defining $ \sigma_{\mu} = (1, \vec{\sigma})$ and $\bar{\sigma}_{\mu} = (1, -\vec{\sigma})$, we then split the four--component
notation into the Weyl formalism as follows,
\be
\psi_{if} = \left ( \matrix{\xi_{\alpha} \cr \bar{\lambda}^{\dot{\alpha}} } \right )_{if} \ \ \ , \ \ \ \ 
\gamma^{\mu} = \left ( \matrix{ 0 \, & \sigma^{\mu} \cr \bar{\sigma}^{\mu} \, & 0} \right ) \ .
\ee
Writing the action (\ref{startaction}) in the Weyl two--component formalism we have, after a partial integration,
\be
S = -i \sum_{f=1}^{N_f} \int d^4 \! x \, \left [ \bar{\lambda}_{jf} \bar{\sigma}^{\mu} \left ( \del_{\mu} \delta^j_i 
+ig A_{\mu}^a (T^a)^j_i \right ) \lambda_f^i + \bar{\xi}^i_f \bar{\sigma}^{\mu} \left ( \del_{\mu} \delta^j_i 
-ig A_{\mu}^a (T^a)^j_i \right ) \xi_{jf} \right ] \ . \label{weylaction}
\ee
Using now the pseudo--reality relation in SU(2),
\be
\sigma_2 \sigma_a \sigma_2 = -\sigma^T_a \ ,
\ee
where $\sigma^T$ denotes the transpose of $\sigma$, the interaction term for $\lambda$ can be written as,
\be 
\bar{\lambda}_{jf} (T^a)^j_i \lambda_{f}^i = \bar{\lambda}_f T_a^T \lambda_f = 
-(\bar{\lambda}_f\sigma_2 ) T_a (\sigma_2 \lambda_f) \ .
\ee
Since $\lambda_f$ transforms as $\bar{{\bf 2}}$ under a gauge transformation,
$\sigma_2 \lambda_f$ transforms as $\xi$, i.e. like {\bf 2}.
Thus, by defining
\begin{equation}
Q_f = \left \{ \begin{array}{ll}
\xi_f  & f=1,\ldots ,{\rm N}_f \\ \sigma_2 \lambda_f  & f={\rm N}_f +1,\ldots ,2{\rm N}_f  \end{array} \right. \ \ \ , \ \ \ 
\bar{Q}_f = \left \{ \begin{array}{ll} \bar{\xi}_f  & f=1,\ldots ,{\rm N}_f \\ 
\bar{\lambda}_f \sigma_2 & f={\rm N}_f +1,\ldots ,2{\rm N}_f  \end{array} \right. \ ,
\end{equation}
the action in equation (\ref{weylaction}) finally becomes,
\be
S=-i \sum_{f=1}^{2N_f} \int d^4 \! x \, \bar{Q}_f \left ( \del_{\mu} -igA_{\mu}^aT^a \right ) \bar{\sigma}^{\mu} Q_f \ .
\label{enlargedaction}
\ee
Hence, the naive symmetry SU($N_f$)$_L \times$SU($N_f$)$_R \times$U(1)$_B$ is enlarged to an SU(2$N_f$) symmetry, 
which becomes manifest
when the action is rewritten as in equation (\ref{enlargedaction}). It should be noted that the U(1)$_B$ symmetry  
becomes a subgroup of SU(2$N_f$), and also that the derivation 
only holds true at a zero chemical potential $\mu$, since at finite $\mu$ the SU(2$N_f$) symmetry
is explicitly broken
to the original SU($N_f$)$_L \times$SU($N_f$)$_R \times$U(1)$_B$ symmetry \cite{correlators}.

Now consider a spontaneous breaking of the  SU(2$N_f$) chiral symmetry \cite{peskin}. Restricting ourselves to a 
Lorentz and gauge invariant condensate bilinear in $Q$, the structure has to be,
\be
\epsilon_{c_1 c_2} \epsilon^{\alpha \beta} Q^{c_1}_{\alpha} Q^{c_2}_{\beta} 
\propto \epsilon^{\alpha \beta} Q^T_{\alpha} \sigma_2 Q_{\beta}\ ,
\ee
where $c_i$ are the color indices.
In order to obtain overall antisymmetry the flavor structure must also be antisymmetric, and
hence the chiral condensate $\Sigma$ must be of the form, 
\be
\Sigma \propto Q_{i c_1 \alpha} Q_{j c_2 \beta } \epsilon^{c_1 c_2} \epsilon^{\alpha \beta} E^{ij} \ ,
\ee
where $E$ is a (2$N_f$)$\times$(2$N_f$) matrix,
\be
E = \left ( \matrix{ {\bf 0} & {\bf 1} \cr - {\bf 1} & {\bf 0 } } \right ) \ ,    
\ee
with ${\bf 0}$ and ${\bf 1}$ $N_f \times N_f$ matrices.
If $\Sigma \neq 0$, the chiral symmetry is spontaneously broken and $\Sigma$ is only invariant under a subgroup of SU(2$N_f$),
specified by the requirement that for any element $H$ in this subgroup,  
\be
Q^TH^TEHQ = Q^TEQ \ . \label{subgroup}
\ee
The set of matrices $H$ satisfying equation (\ref{subgroup}) form the symplectic group Sp(2$N_f$), so for 
a non--zero value of $\Sigma$,
the symmetry breaking pattern is
SU(2$N_f$) $\rightarrow$ Sp(2$N_f$). Since Sp(2$N_f$) contains (2$N_f^2+N_f$) independent parameters, the number of Goldstone
bosons are $(4N_f^2-1) - (2N_f^2+N_f) = 2N_f^2-N_f-1$.

To describe the symmetry breaking and the corresponding low energy degrees of freedom 
we only need to take into account the condensate and the massless fluctuations around it, i.e. the Goldstone bosons.
The coset space is given by SU($2N_f$)/Sp($2N_f$) and   
the effective action can be parametrized by a ($2N_f$)$\times$($2N_f$) complex antisymmetric matrix $\Phi$,
with ($4N_f^2 -2N_f$) independent real--valued parameters,
\be  
\Phi \sim QQ^T \ , \ \ \ \ \ \Phi_{ij} = -\Phi_{ji} \ .
\ee
Under an SU($2N_f$) transformation $M$, $\Phi$ transforms as $\Phi \rightarrow M\Phi M^T$, and 
the antisymmetry of $\Phi$ is obviously maintained.

The most general, renormalizable effective Lagrangian for $\Phi$, invariant under the SU($2N_f$) symmetry is,
\be
L_0 = \frac{1}{2} {\rm Tr}\left [ (\del_{\mu}\Phi^{\dagger} ) (\del^{\mu}\Phi ) \right ]
 - \frac{m^2}{2} {\rm Tr} \left[\Phi^{\dagger} \Phi \right ] -\lambda_1 \left ( {\rm Tr} \left [ \Phi^{\dagger} \Phi \right ]
\right )^2 - \lambda_2 {\rm Tr} \left [ \left (\Phi^{\dagger} \Phi \right )^2 \right ] \ . \label{renormlag}
\ee
If one adds also a vector field, it is possible to enhance the global symmetry SU($2N_f$) to SU($2N_f$)$\times$SU($2N_f$), 
a fact that was explored in \cite{enhanced}. In this article we will, however, only consider the chiral symmetry breaking
and the associated Lagrangian given above.

Clearly, the Lagrangian in equation (\ref{renormlag}) is invariant under U($2N_f$), and not just SU($2N_f$). 
Additional terms that only respect
the SU($2N_f$) symmetry appear when the axial anomaly is taken into account. The corresponding
lowest--dimensional term for an even, antisymmetric $2n\times 2n$ matrix $\Omega$ is the Pfaffian, 
${\rm Pf} (\Omega)$ \cite{smilgaver}, given by
\be
{\rm Pf} (\Omega) = \frac{1}{2^{n} n!} \sum_P (-1)^P \Omega_{i_1 i_2} 
\Omega_{i_3 i_4} \cdots \Omega_{i_{2n-1} i_{2n}}  \,
\ee
where $i_k = 1,\ldots ,2n$, and $P$ denotes a summation over all permutations of $\{ i_1,\ldots ,i_{2n} \}$, 
with $(-1)^P$ the sign of the permutation. The square of ${\rm Pf} (\Omega )$  is simply the determinant of $\Omega$.

The action invariant only under SU($2N_f$) finally becomes,
\be
S = \int d^4 \! x \, \left \{ L_0 - c \left [ {\rm Pf} (\Phi) + {\rm Pf} (\Phi^{\dagger}) \right ] \right \} 
= \int d^4 \! x \, \left \{ L_0 + L_{{\rm Pf.}} \right \} \ , 
\label{effaction}
\ee
where, in four dimensions, the different coupling constants have mass 
dimensions $[\lambda_i]=0$, $[m]=1$ and $[c]=4-N_f$. 

To study the symmetry breaking pattern SU($2N_f$) $\rightarrow$ Sp($2N_f$), $\Phi$ should acquire a non--zero 
expectation value $\Phi_0$ of the form
\be
\Phi_0 = \phi_0 \left ( \matrix{{\bf 0} & {\bf 1} \cr -{\bf 1} & {\bf 0} } \right ) \ , \label{expvalue}
\ee
where ${\bf 1}$ is the $N_f \times N_f$ unit matrix and $\phi_0$ a constant. 
The matrix structure in equation (\ref{expvalue}) clearly respects
the Sp($2N_f$) symmetry, and when $m^2 < 0$ the potential in the action (\ref{effaction})
is minimized for $\phi_0 \neq 0$. Furthermore, stability at large values of 
$\Phi$ requires $\lambda_2 \geq 0$ and $\lambda_1 + \lambda_2 /2N_f \geq 0$.
With this symmetry
breaking term it is straightforward, although rather tedious,
to check explicitly that there are exactly $2N_f^2-N_f-1$ ($2N_f^2-N_f$) 
number of Goldstone bosons for $c\neq 0$ ($c=0$). 

\section{Critical Behavior at Finite Temperature}
\label{III}

In this section we will study the influence of a finite temperature $T$ on the chiral condensate, and the order of the
phase transition to the SU($2N_f$) symmetric phase
as $T$ reaches the critical temperature, $T \rightarrow T_c$. At zero temperature, $m^2 < 0$ and 
we have $\phi_0 \neq 0$,  breaking the SU($2N_f$) symmetry spontaneously to
Sp($2N_f$), as stated in the previous section. We will first assume that $c=0$, so that the
chiral symmetry is enlarged to U($2N_f$), and later discuss possible consequences of a non--zero $c$.

The critical behavior at finite temperature is determined by the infrared fixed points in the coupling constant space
of $\lambda_1$, $\lambda_2$ and the mass parameter. Hence, to study the renormalization--group flow we 
need to calculate the scale dependence for the these parameters. Although it is of course possible
to obtain the $\beta$--functions from a direct calculation of the relevant Feynman diagrams, 
we will here use the simpler method of the effective action. Needless to say, the two different methods are completely
equivalent. 

When the $\beta$--functions are calculated from the effective action they will correspond
to a system in  $(4-\epsilon)$ dimensions, where $\epsilon \ll 1$. 
But since the extension of the timelike direction is only $\beta = 1/T$
(in the imaginary time formalism), the large scale fluctuations in the vicinity of the critical temperature is
governed by an effective action in {\em three} dimensions. Nevertheless, the $\epsilon$--expansion has proved
to be a useful tool to describe critical behavior in three dimensions, where $\epsilon =1$, both
for the chiral symmetry restoration in real QCD
\cite{Robwilczek} but also, and to a great numerical accuracy, 
in several different condensed matter systems (see e.g. \cite{zinnjustin}).

In the effective action, we need to distinguish between the interaction terms with coupling strength
$\lambda_1$ and $\lambda_2$ respectively. To accomplish that, 
the constant $\phi_0$ will be labeled differently for different matrix elements of $(\Phi_0)_{ij}$,
\be
\phi_0 = \left \{ \begin{array}{rl} \phi_a & i=1, \ldots ,(N_f-1) \ , \ \ j= (N_f+1), \ldots ,(2N_f-1) \ . \\
\phi_b & i=N_f \ , j = 2N_f \ . \\ -\phi_a & i=(N_f+1), \ldots ,(2N_f-1) \ , \ \ j= 1, \ldots ,(N_f-1) \ . \\
-\phi_b & i=2N_f \ , j = N_f \ . \end{array} \right. \label{diffphi}
\ee
Using equation (\ref{diffphi}), assuming that $\phi_0$ is real--valued and
expanding the action (\ref{effaction}) to one--loop order around $\Phi_0$ we find,
\be
S_{{\rm 1-loop}} &=& \int d^4 \! x \, \left \{
-\frac{1}{2} \rho^T \left ( \del^2 +m^2 \right ) \rho -\frac{1}{2} \rho^T M_{{\rm int.}} \rho -
m^2 \left [ (N_f -1) \phi_a^2 + \phi_b^2 \right ] \right. \nonumber \\
&-& \left. 4\lambda_1 \left [(N_f -1) \phi_a^2 + \phi_b^2 
\right ]^2 -2\lambda_2 \left [ (N_f -1) \phi_a^4 + \phi_b^4 \right ] \right \} \nonumber \\
&=& \int d^4 \! x \, \left \{
-\frac{1}{2} \rho^T \left ( \del^2 +m^2 \right ) \rho -\frac{1}{2} \rho^T M_{{\rm int.}} \rho -V_0 \right \} \ .
\ee
In the expansion above, we have put all the $N_f(2N_f-1)$ fluctuations of the complex matrix $\Phi$ 
into a real-valued, $2N_f(2N_f-1)$--component vector $\rho$, where $M_{{\rm int.}}$ contains all the nontrivial
terms depending on $\lambda_1$ and $\lambda_2$. In the last line, $V_0$ is nothing but the classical potential, and
by disregarding the different labelings and replacing $\phi_a = \phi_b = \phi_0$, it has a minimum for
\be
\phi_0^2 = \frac{-m^2}{4(2N_f \lambda_1 +\lambda_2)} \ \ \ \ \ \ \ (-m^2 > 0) \ .
\ee

Since $M_{{\rm int.}}$ can be made symmetric, it can be diagonalized by an orthonormal matrix $P$,
\be
M_D = PM_{{\rm int.}} P^T = {\rm diag} (m_1^2, m_2^2,\ldots, m^2_{2N_f(2N_f-1)}) \ ,
\ee
and by changing variables to $\tilde{\rho} = P\rho$ and $\tilde{m}_k^2 = m_k^2 + m^2$, $k=1,\ldots,2N_f(2N_f-1)$,
we find the following Euclidian effective potential after a gaussian integration,
\be
V_{{\rm eff}} = V_0 +\frac{1}{2V_4} \sum_k {\rm LogDet} (-\del^2_{E} + \tilde{m}_k^2) \ , \label{effpotential}
\ee
where $V_4$ is the four--dimensional volume and the subscript $E$ denotes a Euclidean metric $\delta_{\mu \nu} = {\rm diag} (++++)$.

Now, as the temperature $T$ is turned on there is a thermal correction $m_T$ to the zero temperature
mass parameter $m$, and when the temperature reaches the critical temperature $T_c$ we have 
$\phi_0 \rightarrow 0$, or in other words $m^2 + m_{T_c}^2 \rightarrow 0$.
Hence, the critical behavior occurs when the total mass $m^2 + m_T^2$ is 
fine tuned to zero, as usual in critical phenomena. 
Therefore, in the vicinity of the fixed point we can ignore the $m$ dependence in 
the effective potential (\ref{effpotential}), and keep only the remaining
divergent pieces in the second term to calculate
the $\beta$--functions for $\lambda_1$ and $\lambda_2$. We find, using dimensional regularization in $(4-\epsilon)$
dimensions,
\be
V_{{\rm eff}} = V_0 - \frac{1}{32\pi^2 \epsilon} \sum_k m_k^4 \ .
\ee
Finding all the eigenvalues $m_k$ of $M_{{\rm int.}}$, for arbitrary $N_f$, seems a formidable task.
Fortunately, this is not necessary, since
\be
\sum_k m_k^4 = {\rm Tr} M_D^2 = {\rm Tr} M_{{\rm int.}}^2 \ .
\ee
Thus, we can use the matrix $M_{{\rm int.}}$ directly in the effective potential,
\be
V_{{\rm eff}} = V_0 - \frac{1}{32\pi^2 \epsilon} {\rm Tr} M_{{\rm int.}}^2 \ . \label{actionwithM}
\ee
Finally, by writing down the explicit form of $M_{{\rm int.}}$ the $\beta$--functions become,
\be 
\begin{array}{rl}
\beta_1 = \mu \frac{\del \lambda_1}{\del \mu} & = -\epsilon \lambda_1 + \left (\frac{2N_f^2-N_f+4}{\pi^2} \right ) 
\lambda_1^2 + \left (\frac{4N_f-2}{\pi^2} \right ) \lambda_1 \lambda_2 + \left (\frac{3}{2\pi^2} \right ) \lambda_2^2
 \ , \\  &  \\
\beta_1 = \mu \frac{\del \lambda_2}{\del \mu} & = -\epsilon \lambda_2 + \left (\frac{4N_f-5}{2\pi^2} \right )
\lambda_2^2 + \left (\frac{6}{\pi^2} \right ) \lambda_1 \lambda_2 \ , \end{array} \label{betas}
\ee
where $\mu$ is an arbitrary mass scale.

As is well known, an infrared fixed point $\lambda^{\ast}$ corresponds to $\beta_i = 0$ and positive 
eigenvalues of the matrix 
$W_{ij} = (\del \beta_i / \del \lambda_j ) |_{\lambda = \lambda^{\ast}}$, for $i,j=1,2$. 
From equation (\ref{betas}) we find that the only infrared stable point to order $\epsilon$
is given by
\be
\lambda_1^{\ast} = \frac{\pi^2 \epsilon}{2N_f^2 -N_f +4}  \ \ , \ \ \ \ \  \lambda_2^{\ast} = 0 \ ,
\ee
and only for $N_f \leq (1+\sqrt{17})/4  = N_f^{{\rm max}} \simeq 1.28$.
When $N_f > N_f^{{\rm max}}$ the fixed point is infrared unstable along the $\lambda_2$--axis.

Assuming that the predictions from the $\epsilon$--expansion remain valid in three dimensions and in addition that
the theory evolves from a point within the stability region ($\lambda_2 \geq 0$, 
$\lambda_1 + \lambda_2 / (2N_f ) \geq 0$), we
conclude that there is a second order phase transition in two--color QCD with 
$N_f \leq (1+\sqrt{17})/4$ massless flavors, to a phase where the spontaneously broken
U($2N_f$) symmetry is restored and the chiral condensate disappears. 
Physically, the only relevant case is $N_f=1$ where there is just
one independent, complex element in $\Phi$. The phase transition will thus be characterized by the critical exponents of
the O(2) linear $\sigma$--model. For two or more massless flavors, the instability in the direction of $\lambda_2$
should result in a first order phase transition induced by fluctuations \cite{amit}, the analogy in $(4-\epsilon )$
dimensions of the Coleman--Weinberg mechanism \cite{cw}.

Let us now consider the effects of having a non--zero value for $c$ around the critical temperature.
In that case, we have to add to the Lagrangian the term
\be
L_{{\rm Pf.}} = - c \left ( {\rm Pf} (\Phi) + {\rm Pf} (\Phi^{\dagger}) \right ) \ .
\ee
For a single flavor this is a linear term, so $c$ acts like an external magnetic field. As a result,
the symmetry is never really restored,
since for $c \neq 0$ the potential always favors a non--zero expectation value. 
For two flavors, the symmetry group is SU(4)$\sim$O(6) and only half the number 
of fields become massless at the transition point.
The phase transition could therefore be of second order and fall into the O(6) universality class. 

Adding a third flavor, the term proportional to $c$ becomes cubic, which in itself drives the phase transition first
order. For $N_f \geq 4$, the operator is either marginal ($N_f =4$) or irrelevant ($N_f > 4$) 
in four dimensions. In both cases,
the critical behavior is already determined by $\lambda_1$ and $\lambda_2$, so the transition remains 
first order, induced by fluctuations.

\section{conclusions}
\label{IV}

We have presented a study of the chiral condensate in two--color QCD at finite temperature and zero chemical potential.
Based on an effective theory, we find conclusive evidence for a high temperature region where the condensate
disappears. Without the anomaly, there is a second order phase transition for one flavor, and a first order phase transition 
induced by fluctuations when $N_f > (1+\sqrt{17})/4$. 

When the anomaly is taken into account, the situation becomes rather different for $N_f \leq 2$, 
since there is strictly speaking no phase transition for $N_f =1$ and possibly a second order phase 
transition for $N_f=2$.
When $N_f \geq 3$, the phase transition is still first order, but for $N_f=3$ it is not induced by fluctuations.

The order of the phase transition for a small number of flavors is therefore quite sensitive to the value
of the instanton--induced term $c$, a situation rather similar to the one found for the chiral symmetry
in full QCD \cite{Robwilczek}. 

As with three colors, instantons in two--color QCD
become less significant as the temperature is increased,
since the instanton size is cut off by the temperature \cite{finitetrev}. 
Hence, the temperature dependence of $c$ is what really
matters. At $T=0$ the anomaly term should of course be present, but it could happen that $c$ decreases with 
an increasing temperature, and in such a way that
it does not play any significant role at the temperature scale where the   
condensate disappears. In that case, the critical behavior is governed by a restoration to the U($2N_f$) symmetric
phase. The only exception occurs for $N_f=1$, where any relic of the linear term destroys the second order phase transition at $c=0$.  
If, on the other hand, $c$ is not neglectable, the symmetric phase corresponds to a restoration of the SU($2N_f$) symmetry.

Keeping in mind that we have considered a simplified version of QCD, it is somewhat
reassuring to find a behavior reminiscent of the chiral symmetry in full QCD at finite $T$,
in particular since both the original symmetry and the remaining subgroup are rather different
in the two cases. This may indicate that some of the qualitative features of two--color
QCD can indeed be taken over to real QCD, at least with some care.

Another interesting question is how the results derived here are related to the finite temperature behavior of 
two--color QCD at $\mu > 0$.
At zero temperature, the formation of 
a diquark condensate seems to be well established by now, and in the chiral limit
the condensate is nonvanishing at all chemical potentials, which implies that the U(1)$_B$ part
of the overall SU($N_f$)$_L \times$SU($N_f$)$_R \times$U(1)$_B$ symmetry at finite $\mu$ is spontaneously broken
(the symmetry breaking pattern is 
SU($N_f$)$_L \times$SU($N_f$)$_R \times$U(1)$_B \rightarrow$Sp($N_f$)$\times$Sp($N_f$) \cite{correlators}).
Recent lattice simulations at nonzero quark masses $m$ \cite{latticerev} seem to indicate that 
the chiral condensate is nonzero at $\mu = 0$ and then
decreases smoothly as the chemical potential becomes larger, whereas the diquark condensate is zero up to some $\mu_0$ and then increases. Most 
likely, there is a genuine phase transition separating the two regions at some critical chemical potential $\mu_c$, and 
there is presumably also a mixed phase for $\mu_0 \lesssim \mu \lesssim \mu_c$, where none of the condensates vanish. 

Thus, even though the SU($2N_f$) is explicitly broken by both a finite mass $m$ and a finite chemical potential $\mu$, the chiral 
condensate is still nonvanishing for small values of $m$ and $\mu$, as long as $\mu \lesssim \mu_c \sim \sqrt{m}$.
Consequently, if both $m$ and $\mu$ can be treated as small perturbations, 
it is not inconceivable to assume that the chiral phase transition at finite $T$ can be modified to also include the case $m, \, \mu >0$.
For example, a second order phase transition
could be changed into a cross--over region, whereas a first order one presumably would become weaker, but remain. 
By continuity arguments, the chiral phase transition (or crossover)
could also meet the phase transition separating the 
chiral and diquark condensate, possibly at a critical or 
tricritical point. Furthermore, since there is no condensate at all above
the finite $T$ phase transition presented in this letter (at $\mu =0$), this could indicate that there is yet another 
phase transition at finite $T$ and $\mu$, between the superfluid and non--superfluid phases.   

Although the above comments are consistent with the 
lattice results obtained so far, 
it should be emphasized that the remarks are highly speculative at the present stage.
Hopefully, the combined results of future analytical and lattice investigations will be able to clarify the situation.
\vskip 6mm

\begin{center}
{\bf Acknowledgements}
\end{center}
The author thanks R. D. Pisarski for proposing the idea, for interesting discussions and for reading the manuscript.
The author also thanks F. Sannino for kindly sharing his notes on the effective Lagrangian for $\Phi$, and 
the Nuclear Theory group at BNL for their kind hospitality.

\end{document}